\documentclass[runningheads]{llncs}
\usepackage[T1]{fontenc}
\usepackage{booktabs}
\usepackage{multirow}
\usepackage{graphicx}
\usepackage{amsmath}
\usepackage[numbers,sort&compress]{natbib}
\usepackage[
    colorlinks=true,
    linkcolor=blue,
    citecolor=blue,
    urlcolor=blue
]{hyperref}
\renewcommand{\cite}{\citep}

\usepackage{xcolor}
\usepackage{graphicx}
\usepackage{enumitem}

\usepackage{todonotes}
\begin{document}
\title{Does Reasoning Make Search More Fair? \\ Comparing Fairness in Reasoning and Non-Reasoning Rerankers}
\titlerunning{Does Reasoning Make Search More Fair?}
\author{Saron Samuel\inst{1}\orcidID{0009-0002-1827-2023}  \and 
Benjamin Van Durme \inst{1,2}\orcidID{0000-0003-4328-4288}
\and 
Eugene Yang\inst{1,2}\orcidID{0000-0002-0051-1535}}

\authorrunning{S. Samuel et al.}
\institute{Johns Hopkins University \and Human Language Technology Center of Excellence \\
\email{ssamue21@jhu.edu, vandurme@jhu.edu, eugene.yang@jhu.edu}}

\maketitle              %
\vspace{-5mm}
\begin{abstract}
While reasoning rerankers, such as Rank1, have demonstrated strong abilities in improving ranking relevance, it is unclear how they perform on other retrieval qualities such as fairness. 
We conduct the first systematic comparison of fairness between reasoning and non-reasoning rerankers. Using the TREC 2022 Fair Ranking Track dataset, we evaluate six reranking models across multiple retrieval settings and demographic attributes.
Our findings demonstrate reasoning neither improve nor harm fairness compared to non-reasoning approaches. Our fairness metric, Attention-Weighted Rank Fairness (AWRF) remained stable (0.33-0.35) across all models, even as relevance varies substantially (nDCG 0.247-1.000). Demographic breakdown analysis revealed fairness gaps for geographic attributes regardless of model architecture. These results indicate that future work in specializing reasoning models to be aware of fairness attributes could lead to improvements, as current implementations preserve the fairness characteristics of their input ranking.

\end{abstract}
\section{Introduction}

Search systems play a critical role in shaping how we access information since their rankings determine which perspectives gain visibility. 
While surfacing relevant information usually yields better utility in tasks~\cite{gao2024retrievalaugmentedgenerationlargelanguage}, ensuring a comprehensive coverage of different opinions and sources is critical when information systems are used for decision making~\cite{redi2021taxonomyknowledgegapswikimedia}.%
 Recently, as reasoning large language models (LLMs) have shown strong effectiveness in various tasks through test-time reasoning~\cite{deepseek}, rerankers based on these reasoning models can reason about a ranking, generating thoughts and justifications before producing a final ranking. 
These reasoning rerankers, such as Rank1 \cite{rank1}, Qwen3-Reranker \cite{qwen3_embedding_model}, and ReasonRank \cite{reasonrank}, have demonstrated strong improvements in relevance metrics across several benchmarks, such as BEIR~\cite{beir}, NeuCLIR~\cite{lawrie2024overviewtrec2023neuclir} and BRIGHT~\cite{bright}.

While reasoning rerankers may produce more relevant rankings, how they impact fairness in a ranking remains underexplored. Fairness in ranking can be operationalized 
as an equitable distribution of exposure across groups defined by sensitive attributes (such as gender, age, occupation, geography) while still maintaining retrieval effectiveness~\cite{ekstrand2023overviewtrec2022fair}. For example, a search system might overexpose content about men when women are equally relevant.

On one hand, reasoning could help by encouraging the model to deliberate more deeply about context. On the other hand, the reasoning process itself is shaped by pretraining data and internal biases~\cite{lee2025implicitbiaslikepatternsreasoning}. If those biases are reflected in the generated justifications, reasoning may instead amplify unfairness by producing confidently skewed rationales. 

Consider a search query \emph{``basic overview of sailing and types of sailboats.''} Searching over the Fair Ranking 2022 track document collection, Table~\ref{tab:qualitative-comparison} shows, all three rankings consist of relevant documents. BM25 \cite{10.1561/1500000019} retrieval surfaces mainly USA-based articles (7 of 10), while an oracle ranking constructed from relevance annotations achieves better geographic diversity with articles from Sweden, Finland, Israel, UK, Poland, and New Zealand. 
Rank1~\cite{rank1}, a pointwise reasoning reranker, reranks on top of BM25, also outputs a  diverse ranked list with different document origins. 

\begin{table*}[t]
\caption{Qualitative comparison of subject geography between a BM25 initial retrieval, then reranked by Rank1, and the ground truth Oracle rankings for the query ``basic overview of sailing and types of sailboats'' using the Fair Ranking 2022 document collection. These were marked by hand annotation (country level), and closely followed the original annotations (continent level). Sections with -- mean that the subject geography could not be identified.}
\label{tab:qualitative-comparison}
\setlength{\tabcolsep}{1.5mm}
\centering
\resizebox{\textwidth}{!}{
\begin{tabular}{ll|ll|ll}
\toprule
\multicolumn{2}{c|}{\textbf{BM25}} & 
\multicolumn{2}{c|}{\textbf{Rank1}} & 
\multicolumn{2}{c}{\textbf{Oracle Ranking}} \\
\textbf{Title} & \textbf{Origin} & 
\textbf{Title} & \textbf{Origin} & 
\textbf{Title} & \textbf{Origin} \\
\midrule
Outline of sailing & -- & Boating & -- & Seafarer 37 & USA \\
Ultimate 20 & USA & Outline of sailing & -- & Stefan Krook & Sweden \\
Moore 30 & USA & Sailboat & -- & Jali Mäkilä & Finland \\
Catalina 275 Sport & USA & Puffer (dinghy) & USA & Yoav Omer & Israel \\
Ranger 22 & USA & LaserPerformance & UK/USA & Humber Keel & UK \\
Catalina 16.5 & USA & Sandeq & Indonesia & Helsen 22 & USA \\
CS 44 & Canada & MC Scow & USA & Brandworkers Int'l & USA \\
Nonsuch 36 & Canada & Kaep & Palau & David Barnes & New Zealand \\
Hunter 45 DS & USA & Pearson 26 & USA & Joanna Burzyńska & Poland \\
Hunter 33.5 & USA & Inclusion Catamaran & Portugal & Newport 214 & USA \\
\bottomrule
\end{tabular}
}

\end{table*}

To investigate this, we conduct the first systematic comparison of fairness between reasoning and non-reasoning rerankers. 
We want to answer the following research questions:
\begin{itemize}
    \item[\textbf{RQ1}] How does query formulation impact fairness and relevance in both reasoning and non-reasoning rerankers?
    \item[\textbf{RQ2}] Do reasoning rerankers exhibit different fairness characteristics compared to non-reasoning rerankers? 
    \item[\textbf{RQ3}] How does fairness vary across different demographic attributes for reasoning vs non-reasoning rerankers?
\end{itemize}

This work examines exposure-based group fairness, where documents from different demographic groups are expected to receive visibility proportional to their representation among relevant documents and to real-world population statistics. We assess fairness using Attention Weighted Rank Fairness (AWRF), which quantifies how closely the position-weighted exposure distribution across protected groups matches a target fair distribution.

Using the TREC 2022 Fair Ranking Track dataset \cite{ekstrand2023overviewtrec2022fair}, we evaluate models across multiple sensitive attributes: age, alphabetical (topic-based), gender, languages, occupation, popularity, source geography, and subject geography. We experiment with two query settings, original keyword queries, and rewritten queries generated via gpt-4o-mini \cite{openai2024gpt4ocard}. \\

Our contributions are as follows:
\begin{itemize}
    \item We conduct the first systematic comparison of fairness between reasoning and non-reasoning rerankers for information retrieval.
    \item We demonstrate that reasoning capabilities, as currently implemented, neither improve nor harm fairness compared to non-reasoning approaches.
    \item We show that query formulation substantially impacts both relevance and fairness, with natural language queries improving effectiveness for all models.
    \item We find demographic differences in fairness, particularly subject geography attribute consistently showing 10-15\% lower fairness than other attributes across all models and retrieval settings.
\end{itemize}

\section{Background}
\subsection{Reasoning vs. Non-Reasoning Rerankers}

Rerankers refine an initial list of candidate documents (from BM25 \cite{10.1561/1500000019} or dense retrieval like Qwen3 \cite{qwen3_embedding_model}) by re-evaluating their relevance to a given query. Traditional rerankers learn implicit notions of relevance from labeled pairs, while recent approaches integrate explicit reasoning to simulate step-by-step judgment at query time. We distinguish between reasoning and non-reasoning rerankers.

Non-reasoning rerankers predict a scalar relevance score or an ordered list of document id without generating explicit reasoning chains~\cite{nogueira-etal-2020-document, pradeep2023rankzephyreffectiverobustzeroshot}. Reasoning rerankers, such as Rank1~\cite{rank1}, generate intermediate reasoning steps before assigning a relevance score. These models aim to emulate human-like thinking trained with distillation from larger reasoning models such as DeepSeek R1~\cite{rank1, rankk} or explicit preference optimization~\cite{zhuang2025rankr1enhancingreasoningllmbased}. This difference makes reasoning rerankers a good test case for fairness.

\subsection{Fairness in Ranking}
Fairness in ranking has been a subject long studied, from resume ranking for jobs~\cite{10.1145/3696457} to recommendation systems~\cite{burke2017multisidedfairnessrecommendation}. The challenge of balancing relevance with fairness has been approached from multiple perspectives, including individual fairness~\cite{dwork2011fairnessawareness}, group fairness~\cite{Zehlike_2017}, and exposure-based fairness~\cite{Singh_2018,Diaz_2020}.

Early work on fair ranking focused on ensuring that qualified candidates receive appropriate representation in ranked lists. Zehlike et al.~\cite{Zehlike_2017} introduced FA*IR, an algorithm that ensures minimum representation of protected groups. Singh and Joachims~\cite{Singh_2018} proposed fairness of exposure metrics, arguing that items should receive exposure proportional to their relevance. Diaz et al.~\cite{Diaz_2020} extended this work by introducing expected exposure metrics for stochastic ranking policies. Wang et al. ~\cite{wang2024largelanguagemodelsrank} conducted a study on fairness of LLMs as rankers, finding that LLMs can  exhibit demographic biases in ranking decisions.

\subsection{TREC Fair Ranking Track}
The TREC Fair Ranking Track~\cite{ekstrand2023overviewtrec2022fair} introduced a benchmark for studying how retrieval systems can balance relevance and fairness in document exposure. We evaluate static ranked lists using normalized Discounted Cumulative Gain (nDCG) for relevance and Attention-Weighted Rank Fairness (AWRF) ~\cite{10.1145/3308560.3317595, raj2022comparingfairrankingmetrics} for fairness. AWRF measures how closely the exposure of documents across protected groups (gender, occupation, geography, etc.) matches a target distribution that combines real-world population data with the distribution among relevant articles. AWRF compares the actual exposure distribution with a target fair distribution using the Jensen-Shannon divergence, $d_{\text{JS}}$ \cite{menendez1997jensenshannondivergence}.
AWRF is calculated as: 
\begin{equation}
    \text{AWRF}(L) = 1 - d_{\text{JS}}(\mathbf{d}_L, \mathbf{d}_q)
\end{equation}
where $\mathbf{d}_L$ is the normalized, position-weighted exposure distribution across groups in ranking $L$ and $\mathbf{d}_q$ is the target distribution based on group representation in relevant documents and global demographics for query $q$. AWRF ranges from 0 to 1, with 1 indicating perfect fairness. The final system score, $M_1$, combines both AWRF and nDCG as:
\[
M_1(L) = \text{AWRF}(L) \times \text{nDCG}(L)
\]
rewarding systems that maintain both metrics.

The query texts were made from extracted keywords from articles relevant to a WikiProject using KeyBERT \cite{grootendorst2020keybert}. For each WikiProject, keywords were aggregated from relevant articles and manually filtered to form the final query texts. The corpus consisted of articles from English Wikipedia. Relevance was obtained from existing WikiProject page lists. If a WikiProject had tagged an article as being within its scope, that article was considered relevant for queries representing that WikiProject.  The dataset includes multiple fairness categories derived from Wikipedia and Wikidata metadata that are later used to measure AWRF. This includes geographic location, gender for biographical articles, occupation, article age, alphabetical position, article creation date, popularity based on pageviews, and cross-language replication.

This track provides an important pre-reasoning baseline for understanding fairness trade-offs in retrieval systems and serves as the basis for our comparison. With this foundation, we design a set of experiments to investigate how reasoning and non-reasoning rerankers compare.

\section{Experiment Protocol}
We design our set of experiments to address our research questions. 

\begin{itemize}

\item[RQ1] \textbf{Query Formulation.} We evaluate all reranking models with either original keyword queries from Fair Ranking 2022 or rewritten queries. This is done with BM25 as an initial retriever.

\item[RQ2] \textbf{Reasoning vs. Non-Reasoning.} We compare six rerankers. Three are reasoning models (Rank1, Qwen3-Reranker, ReasonRank) and three are non reasoning (MonoT5, RankZephyr, RankLLaMA).

\item[RQ3] \textbf{Demographic Breakdown.} We compute $M_1$ separately for the eight demographic attributes (age, alphabetic, gender, languages, occupation, popularity, source geography, subject geography) across all retrieval settings.

\end{itemize}

\subsection{Initial Retrieval Settings}

Before applying rerankers, we established four initial retrieval settings.\\

\begin{itemize}[topsep=0pt,partopsep=0pt,parsep=0pt]
    \item BM25 with keyword queries
    \item BM25 with rewritten queries
    \item Qwen3-Embedding-8B with rewritten queries
    \item RRF fusion combining BM25 (rewritten) and Qwen3 rankings
\end{itemize}
\vspace{6pt}

Queries are rewritten using gpt-4o-mini with the original query title plus a subset of four randomly selected keywords from the list, which are then used along side a prompt. We rewrite queries because LLM-based rerankers are designed to process
natural language rather than keyword lists. Query rewriting transforms keyword collections ("Sailing, ocean, sailboat, paddle") into a coherent query ("basic overview of sailing and types of sailboats"), better matching how users formulate real search queries. This allows us to evaluate whether query formulation impacts fairness independently of model architecture.

We also constructed an oracle ranking that approximates an upper bound on relevance by reordering documents to achieve a nDCG of 0.9 for the top 500 documents. This is based on the dataset's gold judgments. This oracle setting serves as a reference for how much fairness remains even when retrieval relevance is nearly optimal. The oracle retrieval setting is produced via the TREC 2022 Fair Ranking Track provided script.\footnote{\url{https://github.com/fair-trec/trec2022-fair-public/blob/main/oracle-runs.py}}

\subsection{Rerankers}
For reranking, we select reankers with roughly the same number of parameters to eliminate the effect of different model sizes.

\subsubsection{Listwise Rerankers}

\begin{itemize}
\item RankZephyr-7B-V1 \cite{pradeep2023rankzephyreffectiverobustzeroshot} (refered to as RankZephyr-7B  in tables) is zero-shot listwise reranker built on a 7B Zephyr backbone and instruction finetuned by distilling teacher reorderings from RankGPT3.5 and a smaller set of RankGPT4 outputs.
\item ReasonRank-7B \cite{reasonrank} is a reasoning listwise reranker built on Qwen2.5-7B-Instruct. It is trained through a two-stage framework combining supervised fine-tuning on synthesized reasoning-intensive data and reinforcement learning with multi-view ranking rewards.
\end{itemize}

\subsubsection{Pointwise Rerankers}
\begin{itemize}

\item RankLLaMA-7B \cite{ma2023finetuningllamamultistagetext} is a pointwise reranker fine-tuned from the LLaMA-2-7B architecture for multi-stage text retrieval. The model takes query-document pairs as input and projects the final hidden state of the end-of-sequence token through a linear layer to produce scalar relevance scores. RankLLaMA is trained using contrastive loss with hard negatives sampled from RepLLaMA's top ranking results.

\item MonoT5-base-msmarco-10k \cite{nogueira-etal-2020-document} (refered to as MonoT5-0.3B in tables) is a pointwise reranker based on the T5-base encoder-decoder architecture. It is fine-tuned on the MS MARCO passage ranking dataset for 10,000 steps to predict document relevance as a text generation task. It produces “true” or “false” tokens for each query-document pair and converts the logits of these generated tokens into relevance probabilities for ranking.

\item Qwen3-Reranker-8B \cite{qwen3_embedding_model} is a cross-encoder reranker built on the Qwen3 language model architecture, utilizing 8 billion parameters to assess query-document relevance through joint encoding of query-document pairs. 

\item Rank1-7B \cite{rank1} is a pointwise reasoning reranker distilled from DeepSeek-R1's reasoning traces on MS MARCO, trained via supervised fine-tuning on 635,000 examples of query-passage relevance judgments with step-by-step reasoning chains. The model is built on Qwen 2.5 base models and uses test-time compute to generate explainable reasoning before producing binary relevance predictions.

\end{itemize}

We evaluated these reasoning and non-reasoning rerankers across our four initial retrieval and oracle settings. We rerank the top 500 of each ranking. Note, the rewritten queries and reranker prompts are all neutral, without introducing the concept of fairness.

\subsection{Metrics and Statistical Tests}

In this study, we report nDCG@10 for relevance and AWRF@10 for fairness. 
Instead of evaluating the full ranking, which is what the Fair Ranking track originally proposed, we use a shallow rank cutoff to better evaluate the reranking models.

When testing for \textit{differences} in either nDCG and AWRF, we conduct paired $t$-tests on the query level, where the null hypothesis is that methods are identical ($H_0: x_1 = x_2$). 
Alternatively, when testing for \textit{equivalence}, we use the paired Two One-Sided Tests (TOST)~\citep{schuirmann1987}, where one tests if the difference is greater than the lower bound ($-\delta$) and another tests if it is less than the upper bound ($+\delta$). 
If both tests reject their respective null hypotheses at a certain significance level (typically 0.05), we conclude the methods are statistically equivalent within the tolerance $\delta$ ($H_0: -\delta \le x_1 - x_2 \le \delta$). 
We set $\delta = 0.05$ for all TOSTs conducted in this study.

We apply the Holm-Bonferroni method~\cite{holm-method} for multiple testing correction when rejecting multiple statistical tests in one claim. This applies to both the paired $t$-tests and the paired TOSTs.

\section{Results and Analysis}

\subsection{Relevance and Fairness Comparison between Rerankers}

To answer RQ1, \textit{``how does query formulation impact fairness and relevance in both reasoning and non-reasoning rerankers?''}, we compare performance using keywords versus rewritten queries. We measure relevance using nDCG@10 and fairness using AWRF@10 \footnote{we ran the same evaluation @20 and reached the same conclusion}.

Since LLM-based rerankers are designed to process natural language queries, we first explore the performance differences between using the keywords and rewritten queries. Based on the left two sections in Table~\ref{tab:combined-results}, all rerankers achieve substantially higher nDCG@10 when searching with rewritten queries using BM25 compared to keywords, indicating that our zero-shot rewriting process does not introduce significant query drift.

While almost all rerankers are significantly better in nDCG@10 than the initial BM25 ranking, regardless of the queries, except for RankLLaMA (paired $t$-test with $p<0.05$ and multiple testing correction), with strong relative improvements from the initial ranking.

Interestingly, reranking does not seem to affect the relative group exposure at the top of the ranking, even when we are reranking the top 500 documents from the initial ranking. To further confirm this hypothesis, we conduct TOST and verified that all rerankers demonstrate equivalent AWRF@10 to the initial retrieval ($p<0.05$ with multiple testing correction).

Query formulation, when effective, clearly leads to improved relevance as measured by nDCG@10 for both BM25 and subsequent reranking. However, it does not change fairness, resulting in stationary AWRF@10 scores. We observe these trends on both reasoning and non-reasoning rerankers.

\begin{table*}[t]
\centering
\caption{nDCG@10 and AWRF@10 for BM25, Qwen3-Embedding-8B, and various rerankers across keyword, rewritten, and fusion queries. The (R) and (N) prefix indicates reasoning and non-reasoning models.}
\label{tab:combined-results}
\setlength{\tabcolsep}{0.6mm}
\begin{tabular}{l|cc|cc|cc|cc}
\toprule
{} &  
\multicolumn{2}{c|}{\textbf{Keywords}} &
\multicolumn{6}{c}{\textbf{GPT-4o-mini Rewritten Queries}} \\
{} &
\multicolumn{2}{c|}{\textbf{BM25}} &
\multicolumn{2}{c|}{\textbf{BM25}} &
\multicolumn{2}{c|}{\textbf{Qwen3-8B}} &
\multicolumn{2}{c}{\textbf{Fusion}} \\
& nDCG & AWRF & nDCG & AWRF & nDCG & AWRF & nDCG & AWRF \\
\midrule
Initial Retrieval  & 0.247 & 0.338 & 0.540 & 0.336 & 0.311 & 0.347 & 0.416 & 0.344 \\
\midrule
\multicolumn{9}{l}{\textbf{Listwise Rerankers}} \\
\midrule
(N) RankZephyr-7B     & 0.567 & 0.333 & 0.760 & 0.338 & 0.726 & 0.338 & 0.765 & 0.336 \\
(R) ReasonRank-7B     & 0.455 & 0.337 & 0.712 & 0.338 & 0.680 & 0.341 & 0.747 & 0.339 \\
\midrule
\multicolumn{9}{l}{\textbf{Pointwise Rerankers}} \\
\midrule
(N) RankLLaMA-7B      & 0.271 & 0.333 & 0.526 & 0.337 & 0.656 & 0.340 & 0.588 & 0.337 \\
(N) MonoT5-0.3B       & 0.391 & 0.338 & 0.721 & 0.343 & 0.717 & 0.341 & 0.713 & 0.342 \\
(R) Qwen3-Reranker-8B & 0.373 & 0.338 & 0.617 & 0.337 & 0.503 & 0.342 & 0.605 & 0.343 \\
(R) Rank1-7B          & 0.503 & 0.333 & 0.721 & 0.344 & 0.693 & 0.336 & 0.763 & 0.343 \\
\bottomrule
\end{tabular}

\end{table*}

Summarized in the right parts of Table~\ref{tab:combined-results}, the Qwen3 and the fusion initial rankings both have a lower nDCG@10 compared to the BM25 using the rewritten queries (0.540). 
Again, all rerankers demonstrate equivalent group exposure at the top of the ranks from the initial retrieval results measured by AWRF@10 (TOST with $p<0.05$ and multiple testing correction). 
Among the pointwise rerankers, MonoT5, while being the smallest, performs the strongest in nDCG@10 (0.717) when reranking Qwen3's initial ranking, but is slightly worse than Rank1 on reranking fusion. 
Pointwise rerankers all reduce the AWRF score after reranking compared to the initial Qwen3 and fusion retrieval results.

Again, all rerankers demonstrate equivalent AWRF@10 to their respective initial retrieval results (TOST with $p<0.05$ and multiple testing correction). Among the pointwise rerankers, MonoT5, while being the smallest model, achieves the strongest nDCG@10 (0.717) when reranking Qwen3's initial ranking, but performs slightly worse than Rank1 on reranking fusion results (0.713 vs. 0.763). Pointwise rerankers all show lower AWRF@10 after reranking compared to the initial Qwen3 and fusion retrieval results. While the signals are relatively weak, this suggests these pointwise rerankers only consider relevance in the scores they produce, which aligns with their training objectives.

Across all three initial rankings using rewritten queries (BM25, Qwen3, and fusion), listwise rerankers consistently provide stronger effectiveness in nDCG@10, with the non-reasoning RankZephyr being slightly better than the reasoning counterpart ReasonRank. Although still lower than the initial retrieval AWRF@10, ReasonRank shows slightly numerically better AWRF@10 when reranking Qwen3 (0.341) and fusion (0.339) compared to RankZephyr (0.338 and 0.336), indicating that the reasoning model may begin considering document diversity when documents are equally relevant. However, while capable of comparing across documents and considering aspects beyond just relevance signals, these two listwise rerankers do not effectively consider other signals, even when documents are equally relevant, resulting in lower AWRF@10 than the initial ranking.

To answer RQ2 (\textit{``Do reasoning-based rerankers exhibit different fairness characteristics compared to non-reasoning rerankers?''}), we examine fairness patterns in the main results, then isolate relevance and fairness factors using an oracle ranking experiment.

The difference in AWRF@10 between initial retrieval and reranked results can be attributed to a combination of two factors: (1) inability to rank relevant documents to the top, and (2) inability to identify alternative relevant documents with different demographics. Comparing reasoning versus non-reasoning rerankers across all three initial rankings (BM25, Qwen3, fusion) in Table~\ref{tab:combined-results}, we observe minimal differences in AWRF@10. For example, when reranking fusion results, ReasonRank shows AWRF@10=0.339 versus RankZephyr's 0.336, numerically similar values, and both lower than the initial retrieval AWRF@10 of 0.344.

To isolate whether rerankers can consider fairness when relevance is controlled, we construct an oracle ranking of 500 documents that achieves full nDCG of 0.9 (with nDCG@10 of 0.886) and rerank with all tested rerankers. Summarized in Table~\ref{tab:oracle}, all rerankers achieve near-perfect nDCG@10 except for RankLLaMA. Since there are generally more than 10 relevant documents in each query, these results demonstrate that rerankers successfully place almost only relevant documents in the top 10 ranking.

With near-perfect relevance achieved by all rerankers (nDCG@10 $\geq$ 0.923 except RankLLaMA), we can isolate the fairness behavior. Despite all rerankers demonstrating lower or equivalent AWRF@10 compared to the oracle initial retrieval (0.352), we observe weak trends in different directions. Pointwise reasoning models show slightly higher AWRF@10 (Qwen3-Reranker: 0.350, Rank1: 0.348) compared to non-reasoning pointwise models (MonoT5: 0.345, RankLLaMA: 0.345). This implies that when documents are all equally relevant, the pointwise scores produced by reasoning models may have started to convey information beyond pure relevance.

However, for the two listwise rerankers, the trend reverses, RankZephyr shows slightly higher AWRF@10 (0.353) compared to ReasonRank (0.350). We hypothesize that this may be due to the ranking instruction tuning that explicitly pushes the model to consider only relevance signals. Since reasoning listwise models are more capable of following instructions, the training instruction may be amplified in the final reranking results.

There is no strong evidence that reasoning-based rerankers exhibit different fairness characteristics compared to non-reasoning rerankers. While we observe weak trends in the oracle experiment, with pointwise reasoning models showing slightly better AWRF@10 but listwise reasoning models showing slightly worse AWRF@10, these patterns require further investigation to confirm.

This equivalence across architectures can be explained by several factors. First, 
none of the rerankers were explicitly trained to optimize for fairness, they were 
trained solely on relevance judgments from datasets like MS MARCO. Without 
fairness-aware training objectives or prompts, rerankers lack incentive to 
consider demographic attributes. Second, demographic information (especially 
geography) is often not explicitly mentioned in document text, making it difficult 
for rerankers to condition on these attributes even implicitly. Third, rerankers 
operate on fixed candidate pools from initial retrieval. So if the initial ranking 
has limited demographic diversity in the top 500 documents, reranking cannot 
introduce diversity not present in the pool. The reasoning process in models like 
Rank1 focuses on query-document relevance matching rather than demographic 
considerations, explaining why reasoning provides no fairness advantage over 
non-reasoning approaches under current training paradigms.

\begin{table*}[t]
\centering
\caption{Oracle results for all rerankers and retrieval baselines.}
\label{tab:oracle}
\setlength{\tabcolsep}{3mm}
\begin{tabular}{l|cc}
\toprule
{} & nDCG & AWRF \\
\midrule
Oracle Initial Retrieval & 0.886 & 0.352 \\
\midrule
\multicolumn{3}{l}{\textbf{Listwise Rerankers}} \\
\midrule
(N) RankZephyr-7B   & 0.972 & 0.353 \\
(R) ReasonRank-7B   & 1.000 & 0.350 \\
\midrule
\multicolumn{3}{l}{\textbf{Pointwise Rerankers}} \\
\midrule
(N) RankLLaMA-7B    & 0.825 & 0.345 \\
(N) MonoT5-0.3B     & 1.000 & 0.345 \\
(R) Qwen3-Reranker  & 0.923 & 0.350 \\
(R) Rank1-7B        & 0.996 & 0.348 \\
\bottomrule
\end{tabular}
\end{table*}

\subsection{Some Fairness Attributes are Disproportionately Represented}

Table \ref{tab:oracle-attr} provides attribute-level breakdowns across the eight sensitive attributes on $M_1$, which is the official metric used in the 2022 Fair Ranking track. The nDCG@10 values are close to 1.0 as we used the Oracle rankings, so $M_1$@10 is essentially AWRF@10. 
Despite only showing the results of reranking the oracle initial rankings, we observe very similar trends in reranking other initial rankings. 
However, since reranking the oracle initial rankings provides more candidate relevant documents for the rerankers to choose from, it provides cleaner experimental results to analyze. 

Across all retrieval settings, certain attributes consistently achieved higher fairness scores than others. Languages, gender, and age attributes typically scored highest across all rerankers and initial rankings. 
The alphabetical bias (the attribute of biasing toward alphabetical order since it is likely the default processing order of documents) is particularly interesting, where most recent LLM-based rerankers are trying to combat the positional biases~\cite{li2024split}. 
While the oracle initial retrieval has roughly the same AWRF on gender and alphabets, all rerankers exhibit substantially higher AWRF on alphabets than on gender. 

The attributes that are systematically lower in AWRF, such as geographical locations, are ones that are less likely to be represented in the text, which is what the rerankers only considered. 
If these attributes are crucial in the applications but are not incorporated into the retrieval pipeline, it is unlikely that the models will provide a fair ranking on these attributes. 
This indicates a fundamental limitation that certain demographic groups are underrepresented even among highly relevant documents, since this information is generally harder to capture in the text. 

Comparing different rerankers, reasoning rerankers are generally better than the non-reasoning counterparts with the exception of MonoT5, which is comparable with Rank1. 

So to answer RQ3: \textit{How does fairness vary across different demographic attributes for reasoning vs non-reasoning rerankers}, there are strong differences between the attributes for all rerankers, likely due to the availability of the information in text. However, there are weaker differences between reasoning and non-reasnoing rerankers, with reasnoing ones being slightly more fair.

\begin{table}[t]
\centering
\caption{Overall and attribute $M_1$ of each reranker. 
Since AWRF is a distributional distance measurement, overall $M_1$ is not an average over each attribute. 
}
\label{tab:oracle-attr}
\setlength{\tabcolsep}{3pt} %
\resizebox{\textwidth}{!}{
\begin{tabular}{l|c|cccccccc}
\toprule

\textbf{Model}    & $M_1$   & langs & gender & alpha &   age &   pop &   occ & src-geo & sub-geo\\
\midrule
Oracle Retrieval    & 0.312   & 0.873 &  0.868 & 0.867 & 0.864 & 0.866 & 0.854 &  0.797  &  0.761 \\
\midrule
\multicolumn{10}{l}{\textbf{Listwise Rerankers on oracle}} \\
\midrule
(N) RankZephyr-7B   & 0.343  & 0.921 &  0.901 & 0.926 & 0.906 & 0.897 & 0.893 &  0.851  &  0.801 \\
(R) ReasonRank-7B   & 0.350   & 0.941 &  0.926 & 0.937 & 0.930 & 0.913 & 0.916 &  0.864  &  0.820 \\
\midrule
\multicolumn{10}{l}{\textbf{Pointwise Rerankers on oracle}} \\
\midrule
(N) RankLLaMA-7B    & 0.287   & 0.816 &  0.808 & 0.814 & 0.813 & 0.804 & 0.797 &  0.758  &  0.722 \\
(N) MonoT5-0.3B     & 0.345	   & 0.959 &  0.901 & 0.952 & 0.944 & 0.936 & 0.891 &  0.890  &  0.827 \\
(R) Qwen3-Reranker  & 0.323  & 0.905 &  0.891 & 0.900 & 0.898 & 0.896 & 0.884 &  0.836  &  0.800 \\
(R) Rank1-7B        & 0.346	   & 0.952 &  0.936 & 0.968 & 0.937 & 0.915 & 0.924 &  0.887  &  0.829 \\

\bottomrule
\end{tabular}
}

\end{table}

\section{Conclusion and Future Work}

This work presents the first systematic comparison of fairness between reasoning and non-reasoning rerankers across multiple retrieval settings and demographic attributes. Using the TREC 2022 Fair Ranking Track, we evaluated six reranking models across four retrieval configurations and 1 oracle setting: BM25 with keyword queries, BM25 with rewritten queries, Qwen3-8B dense retrieval, BM25+Qwen3-8B fusion, and oracle procduced ranking targeting nDCG = 0.9.

Our primary finding is that reasoning capabilities, as currently implemented in LLM-based rerankers do not improve or harm fairness compared to non-reasoning approaches. 
Attribute-level analysis revealed interesting differences between demographics. Geographic attributes, particularly subject geography, consistently showed the lowest fairness scores. Even in the oracle setting with near-perfect relevance, subject geography peaked at 0.829, compared to 0.968 for alphabetic attributes and 0.959 for languages. These patterns held for both reasoning and non-reasoning rerankers.

Improving search fairness requires interventions beyond reranking such as further diversifying document collections, auditing representational gaps, and designing retrieval strategies that actively surface underrepresented perspectives. Natural language query understanding deserves continued investment, as query formulation influences both relevance and fairness more than reranking architecture.

Limitations include our focus on exposure-based fairness (AWRF), which may not capture other fairness attributes such as calibration or intersectional fairness. We do not examine the content of reasoning justifications, which could reveal subtle biases not captured by ranking metrics. Our evaluation is confined to English-language documents and specific demographic attributes.

Our evaluation uses AWRF with TREC 2022's specific target distribution formulation, 
which equally weights empirical relevant-document demographics and world population 
statistics. The stability of AWRF across rerankers might not hold under alternative 
fairness definitions that place different weights on these components or that 
enforce stricter demographic parity constraints. 

Future work should extend fairness analysis to other contexts, examine whether reasoning can be steered toward more equitable outcomes, and develop methods for improving collection diversity alongside algorithmic fairness. Reasoning rerankers offer strong relevance improvements without fairness degradation but cannot overcome representation gaps in the information ecosystems they operate within. 

\subsection{Theory of Change}
When search systems underrepresent certain demographic groups, they reinforce information inequalities. As LLM based reasoning rerankers become increasingly deployed in production systems, understanding how they impact fairness is important to prevent amplifying demographic biases at scale. Our work compares fairness outcomes between reasoning and non-reasoning rerankers across multiple retrieval settings and demographic attributes.

For this work to achieve its desired impact, several preconditions must hold. Fairness metrics like AWRF must become standard in retrieval evaluation pipelines since reranking for relevance alone cannot solve fairness. Our oracle experiments show that even near perfect relevance prediction does not also achieve significantly higher fairness. Meaningful progress requires addressing upstream issues such as diversifying content sources and implementing retrieval strategies that actively seek a wider coverage of aspects/perspectives.

Potential negative externalities include false complacency. Our findings might be misinterpreted as evidence that modern rerankers are ``fair enough,'' discouraging ongoing auditing when ``not making things worse'' differs from ``making things fair.'' Overreliance on AWRF as the sole fairness metric risks overlooking other attributes like quality of representation or intersectional fairness. 

\subsubsection*{Disclosure of Interests}
The authors have no competing interests to declare that are relevant to the content of this article.

\bibliographystyle{splncs04nat}

\bibliography{biblio}

\end{document}